\documentclass[a4paper,10pt]{article}
\usepackage{amsfonts}
\usepackage{amsmath}
\usepackage{amssymb}
\usepackage{graphicx}
\hoffset= - 1.0in         

\begin{document}
\hfill ITEP/TH-48/09

\vspace{2cm} \centerline{\Large{ Notes on Chern-Simons Theory in
the Temporal Gauge}} \vspace{1cm}

 \centerline{  \textit{Andrey Smirnov} {\footnote{E-mail: asmirnov@itep.ru}} }

\vspace{1cm}

\centerline{\textbf{Abstract}}

 We analyze the
perturbative series expansion of vacuum expectation values (vevs)
for Wilson loop operators in Chern-Simons (CS) gauge theory in the
temporal gauge $A_{0}=0$. Following  J. Labastida and E. P\'erez
we introduce the notion of the kernels of knot polynomial
invariants - the (non-invariant) vevs of the Wilson loops, arising
from CS theory in the temporal gauge. A method for exact
calculation of the kernels of knot polynomial invariants is
presented.
 \vspace{1cm}
\tableofcontents
\section{Introduction}
In 1988, Edward Witten showed the connection between Chern-Simons
field  theory and the theory of knots in three-dimensional space
\cite{Witten}. Since that time the knot theory has been
intensively studied using the standard quantum field theory
methods. Different approaches inherent in quantum field theory
established important connections between different types of knot
invariants and provided a lot of new constructions for them.\\
\indent The CS theory has been studied in different ways. Using
series of non-perturbative methods Edward Witten in his original
paper proved the equivalence of vacuum expectation values for
Wilson lines operators and known polynomial knot invariants
\cite{jones}-\cite{Kauff}. Perturbative studies performed in
series of paper \cite{BarNtan}-\cite{Gukov} established the
connection between the coefficients of perturbative series
expansion of vevs for Wilson loops and finite type (Vassiliev)
knot invariants \cite{Vass1, Vass2}.\\
\indent Perturbative series expansion has been studied for
different gauge fixing. The study of CS theory in the covariant
Landau gauge performed in \cite{BarNtan,Guad} showed the
equivalence of coefficients for the perturbative series and the
integral representation for invariants by Bott and Taubes
\cite{Bot}. The Kontsevich integral representation for Vassiliev
invariants turned out to correspond to the coefficients of vevs of
Wilson lines operators in the light-cone gauge \cite{ Lab3,Kon}.\\
\indent In this paper we concentrate our attention on the study of
 CS theory in the temporal gauge.
 Different aspects of this gauge were
 discussed in \cite{Morozov, Lab2,BF}.  In the temporal gauge $A_{0}=0$ the cubic term of the CS
 action disappears and the main ingredient for perturbative
 calculations is the gauge propagator. As it was noted in
 \cite{Lab2} the calculation of the gauge propagator in non-covariant gauges is
 plagued with ambiguities which should be solved by demanding
 additional properties for the correlation functions of gauge fields.
 As it was shown in \cite{Lab3} in non-covariant gauges we need to
 introduce some additional term to the propagator in order to
 obtain the knot invariants. Unfortunately, at present the
 correction term for the propagator in the temporal gauge is not
 known yet. In this paper we work with the propagator (\ref{Prop})
 without the correction term what leads to non-invariant quantities for vevs
 of the Wilson loop operators. On the other hand side as it was
 argued in \cite{Lab2} the propagator (\ref{Prop}) contains enough information
 about a knot to reconstruct the full knot invariants in any
 order. Following the definitions of \cite{Lab2} we call these
 non-invariant vevs of Wilson loops \textit{the kernels of polynomial
 invariants}.\\
 \indent The aim of this paper is to present the method for calculation
 of the kernels of the knot polynomial invariants in a very simple
 way. We show that there exists an operator $X: R_{1}\otimes R_{2} \rightarrow
 R_{1}\otimes R_{2}$, where $R_{1}$ and $R_{2}$ are some representations of gauge group of CS
 action, such that the kernel of polynomial invariant for a knot with $m$
intersection points can be obtained by appropriate contraction of
indexes for $m$ copies of the operator $X$.\\
\indent The structure of the article is as follows: in section $2$
we discuss the main ingredients of perturbative calculations in CS
theory for the temporal gauge and explain in details the
calculation of vevs for Wilson loops  in the simplest abelian
case. In section $3$ we discuss the geometrical Labastida-P\'erez
formula for perturbative series expansion in non-abelian case and
present some results of explicit calculations with this formula
for the gauge group $GL(N)$. In section $4$ we introduce a notion
of the intersection operator, we calculate explicitly this
operator for the case of $GL(2)$ gauge group. As an example, using
this operator we calculate the exact answer for kernel of
polynomial invariant for the trefoil knot in $GL(2)$ case.
\section{Chern-Simons theory in the temporal gauge}
\subsection{The propagator of the gauge fields in the temporal gauge}
Let $G$ be a semi-simple Lie group and
$A=A_{\mu}^{a}(x)dx^{\mu}F^{a}$ be a $G$-connection on
$\mathbb{R}^{3}$, where $F^{a}$ are the generators of Lie algebra
of $G$ in the fundamental representation. The Chern-Simons field
theory is defined by the following action:
\begin{equation}
\label{CS} S[A]=\int\limits_{\mathbb{R}^3} tr \left(\,A\wedge
dA+g\,\dfrac{2}{3}\,A \wedge A \wedge A \,\right)
\end{equation}
where $tr$ denotes the trace over the fundamental representation
of $G$ and $g$ is some parameter.\\
The CS theory is a particular example of a topological field
theory \cite{BF}, that implies that all the observables in this
theory are some topological invariants. In this way, the CS theory
is a natural tool for study of three-dimensional topology, for
example the topology of three-dimensional knots. Indeed, let us
consider the following gauge invariant operator associated with a
knot $c \in \mathbb{R}^{3}$:
\begin{equation}
W_{R}(c)=tr\left(P\exp\left(g\oint\limits_{c}A_{\mu}^{a}(x)\,R^{a}\,dx^{\mu}\right)
\right),
\end{equation}
this operator is just a Wilson loop associated with the knot $c$
or a trace of the holonomy of $A$ along the path $c$. The index
$R$ means that one-form $A$ takes values in representation $R$ of
$G$. The natural class of the knot invariants provided by the CS
theory is the vacuum expectation values of these operators:
\begin{equation}
\label{VEV} <W_{R}(c)>=\int DA \exp(-S[A])\, W_{R}(c)
\end{equation}
In \cite{Witten} E.Witten using non-perturbative methods showed
that these invariants are in fact the well known polynomial knot
invariants with the argument $t=\exp(g^2)$. To compute (\ref{VEV})
using the standard perturbative methods (with respect to the
parameter $g$)  we need to chose a gauge fixing condition to make
the associated functional integral well defined. Different choices
of a gauge fixing lead to different descriptions of the same
polynomial invariants \cite{Lab2,Lab3,BarNtan}. The aim of this
paper is to study the structure of the perturbative expansion of
these polynomials in the so called temporal gauge. In the temporal
gauge the condition imposed on the field $A$ is:
$$
A_{0}=0
$$
In this case the cubic term of the action (\ref{CS}) disappears
and we arrive to a free field theory with the following action:
\begin{equation}
S[A]|_{A_{0}=0}=\int\limits_{\mathbb{R}^{3}} d^{3}x\,A^{a}_{\mu}\,
(\,\delta^{a\,b}\,\epsilon^{\mu\,0\,\nu}\,\dfrac{\partial}{\partial
x_{0}} \,)\,A^{b}_{\nu},\ \ \ \ \delta^{a\,b}=tr(F^{a}\,F^{b})
\end{equation}
We have the following equation for the Green function:
$$
\delta^{a\,b}\,\epsilon^{\mu\,0\,\nu}\,\partial_{0}\,
\triangle^{b\,c,\,\nu\,\eta}(x)=\delta^{a\,c}\,\delta^{\mu\,\eta}\,\delta(x)
$$
for the solution of this equation we have (for careful derivation
of this formula see \cite{Lab2}):
\begin{equation}
\label{Green}
\triangle^{b\,c,\,\nu\,\eta}(x)=-\epsilon^{\nu\,0\,\eta}\,\delta^{b\,c}\,\delta(x_{1})\,\delta(x_{2})\,
 \dfrac{1}{2}\textrm{sign}(x_{0})+ f^{b\,c\,\nu\,\eta}(x_{1},x_{2}) ,
\end{equation}
where $f^{b\,c\,\nu\,\eta}(x_{1},x_{2})$ is an arbitrary tensor
that does not depend on $x_{0}$. Particular choice of
$f^{b\,c\,\nu\,\eta}(x_{1},x_{2})$ depends on a prescription, some
examples of different prescriptions can be found in
\cite{Lab2,BF}. In this work we will restrict our attention to the
case $f^{b\,c\,\nu\,\eta}(x_{1},x_{2})=0$. In this case one
obtains the following components of the propagator:
\begin{small}
\begin{equation}
\label{Prop}
 <\,A_{0}^{a}(x),\,A_{\mu}^{b}(y)\,>=0, \mu=0,1,2; \ \ \
<\,A_{\mu}^{a}(x),\,A^{b}_{\nu}(y)\,>=\epsilon^{\mu\,\nu}\,\delta^{a\,b}\,\delta(x_{1}-y_{1})\,\delta(x_{2}-y_{2})\,
\dfrac{1}{2}\textrm{sign}(x_{0}-y_{0}), \ \ \mu,\nu=1,2
\end{equation}
\end{small}
As we will see below, perturbation theory based on the propagator
(\ref{Prop}) leads to the quantities for $<W_{R}(c)>$ that do not
coincide with the known knot invariants obtained using
non-perturbative methods and depend on a projection of the knot to
the two-dimensional $(x_{1},x_{2})$ plane. As it was conjectured
in \cite{Lab2,Lab3} this deviation arises because we do not take
into account the prescription depending term in (\ref{Green}).\\
Despite the vacuum expectation values of Wilson lines in this
description are not knot invariants, they contain a lot of
information about knots and as it was conjectured in \cite{Lab2}
they can be used for a reconstruction of the full invariants.
Following the notation of \cite{Lab2} we will denote these
quantities by $< \hat W_{R }(c)>$ and call them \textit{kernels of
polynomial invariants}. The aim of this paper is to present the
method that allows us to find exact expressions for the kernels of
the polynomial invariants for every knot and for every
representation in a very simple and elegant way.
\subsection{The Abelian case as a basic example}
Let us analyze in details the structure of the perturbative series
expansion with propagator (\ref{Prop}) in the Abelian case. In
this case for the propagator we have:
\begin{equation}
\Pi_{\mu\,\nu}(x-y)=<\,A_{\mu}(x),\,A_{\nu}(y)\,>=\dfrac{1}{2}\epsilon^{\mu\,\nu}\,\delta(x_{1}-y_{1})\,\delta(x_{2}-y_{2})\,\textrm{sing}(x_{0}-y_{0}),
\ \ \mu,\nu=1,2,\ \ \ A_{0}=0
\end{equation}
The Wilson lines are presented by a "simple" exponents:
\begin{equation}
W(c)=\exp\left(g\,\oint\limits_{c}A_{\mu}(x)\,dx^{\mu}\right)
\end{equation}
We omit the label of representation because in the Abelian case
the Wilson loop operators for different representations differ by
a constant. For the perturbative expansion we get:
$$
<\hat W(c) >\,=\,<\sum_{n=0}^{\infty} \dfrac{1}{n!}\left(
g\,\oint\limits_{c}A_{\mu}(x)\,dx^{\mu} \right)^{n}>
=\,\sum_{n=0}^{\infty}
\dfrac{(2n-1)!!}{2n!}g^{2n}\left(\int_{c}\int_{c}
dx_{\mu}\,dx_{\nu}<A_{\mu}(x),A_{\nu}(x)>\right)^{2n}
$$
and after the summation we obtain:
\begin{equation}
\label{AbEx} <\hat W(c)>\,=\exp\left(g^2\,L_{c}\right)
\end{equation}
where:
$$
L_{c}=\dfrac{1}{2}\int_{c}\int_{c}
dx_{\mu}\,dy_{\nu}<A_{\mu}(x),A_{\nu}(y)>\,=\dfrac{1}{2}
\int_{c}\int_{c}dx^{\mu}\,dy^{\nu}\dfrac{1}{2}\epsilon^{\mu\,\nu}\,\delta(x_{1}-y_{1})\,\delta(x_{2}-y_{2})\,\textrm{sign}(x_{0}-y_{0})
$$
Let us parametrize  the knot by a parameter $t$ running from $0$
to $1$, then we can rewrite the last integral in the following
form:
$$
L_{c}=\dfrac{1}{4} \int_{0}^{1}\int_{0}^{1} dt_{1} dt_{2}
\left(\dfrac{dx_{1}}{dt_{1}}\,\dfrac{dy_{2}}{dt_{2}}-
\dfrac{dx_{2}}{dt_{1}}\,\dfrac{dy_{1}}{dt_{2}}\right)\delta(x_{1}(t_{1})-y_{1}(t_{2}))\,\delta(x_{2}(t_{1})-y_{2}(t_{2}))\,\textrm{sign}(x_{0}(t_{1})-
y_{0}(t_{2}))
$$
To perform the integration we need to solve the following
equations:
\begin{equation}
\label{Last} \left\{\begin{array}{c}
x_{1}( t_{1} )-y_{1}(t_{2})=0\\
\\
x_{2}( t_{1} )-y_{2}(t_{2})=0
\end{array}\right.
\end{equation}
The solutions of these equations are the self-intersection points
of two dimensional curve $(x_{1}(t),x_{2}(t))$ which is the
projection of the knot $c$ on the plane $(x_{1},x_{2})$. Let us
denote by $t_{1}^{k}<t_{2}^{k}$ the values of the parameter $t$ in
the intersection points, then the two-dimensional delta function
in the integral can be represented in the form:
$$
\delta(x_{1}(t_{1})-y_{1}(t_{2}))\,\delta(x_{2}(t_{1})-y_{2}(t_{2}))=\sum\limits_{k}
\dfrac{\left(
\delta(t_{1}-t_{1}^k)\delta(t_{2}-t_{2}^k)+\delta(t_{1}-t_{2}^k)\delta(t_{2}-t_{1}^k)
\right)}{|\dfrac{dx_{1}}{dt_{1}}\,\dfrac{dy_{2}}{dt_{2}}-
\dfrac{dx_{2}}{dt_{1}}\,\dfrac{dy_{1}}{dt_{2}}|}
$$
Substituting this expression into (\ref{Last}) and integrating
over $t_{1}$ and $t_{2}$ we arrive to the following simple
expression:
\begin{equation}
L_{c}=\sum_{k} \epsilon_{k}
\end{equation}
where the quantities $\epsilon_{k}$ are the "sings" of the
intersection points. They can take values $\pm 1$ and are defined
in the following way:
\begin{equation}
\label{signs}
\epsilon_{k}=\dfrac{\dfrac{dx_{1}}{dt_{1}}(t_{1}^{k})\,\dfrac{dy_{2}}{dt_{2}}(t_{2}^{k})-
\dfrac{dx_{2}}{dt_{1}}(t_{1}^{k})\,\dfrac{dy_{1}}{dt_{2}}(t_{2}^{k})}{|\dfrac{dx_{1}}{dt_{1}}(t_{1}^{k})\,\dfrac{dy_{2}}{dt_{2}}(t_{2}^{k})-
\dfrac{dx_{2}}{dt_{1}}(t_{1}^{k})\,\dfrac{dy_{1}}{dt_{2}}(t_{2}^{k})|}\,\textrm{sign}(x_{0}(t_{1}^{k})-
y_{0}(t_{2}^{k}))
\end{equation}
Note that the expectation value (\ref{AbEx}) for the knot can be
expressed in terms of product over the intersection points:
\begin{equation}
\label{AbAns} <\hat W(c)>\,=\prod\limits_{k}
\exp(g^2\,\epsilon_{k})
\end{equation}
We see that the answer for vev in the abelian case has a form of a
product over the intersection points of some quantity
$\exp(g^2\,\epsilon_{k})$ that depends only on the intersection
point. As we will see below in the non-abelian case this formula
generalizes to the contraction of some tensors corresponding to
intersection points.
\section{The Non-Abelian Case}
\subsection{Labastida-P\'erez formula}
\begin{figure}[h]
\begin{center}
\includegraphics[scale=0.5]{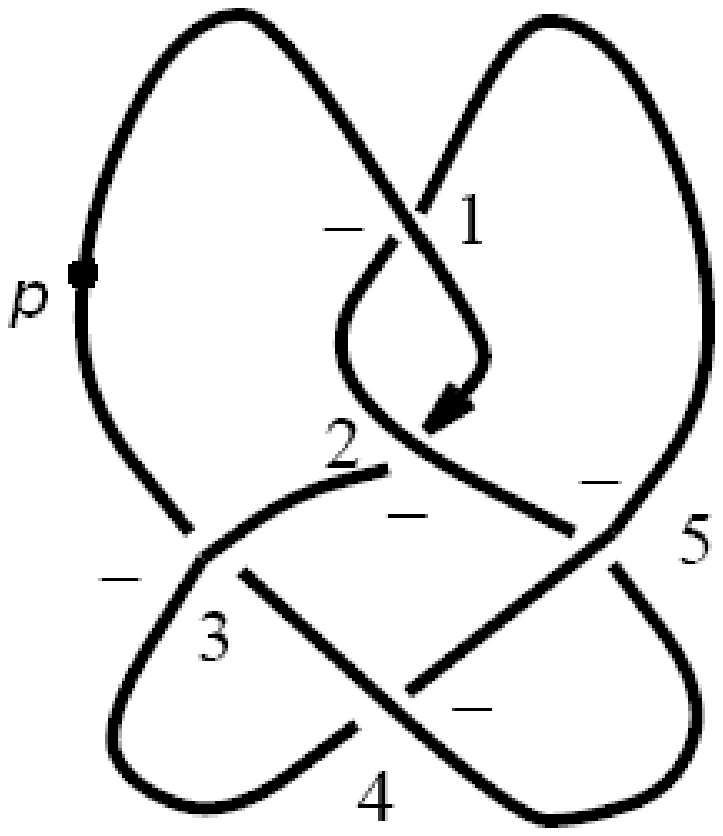}
\caption{ \label{labasgraph}}
\end{center}
\end{figure}

\begin{figure}[h]
\begin{center}
\includegraphics[scale=0.5,angle=-90]{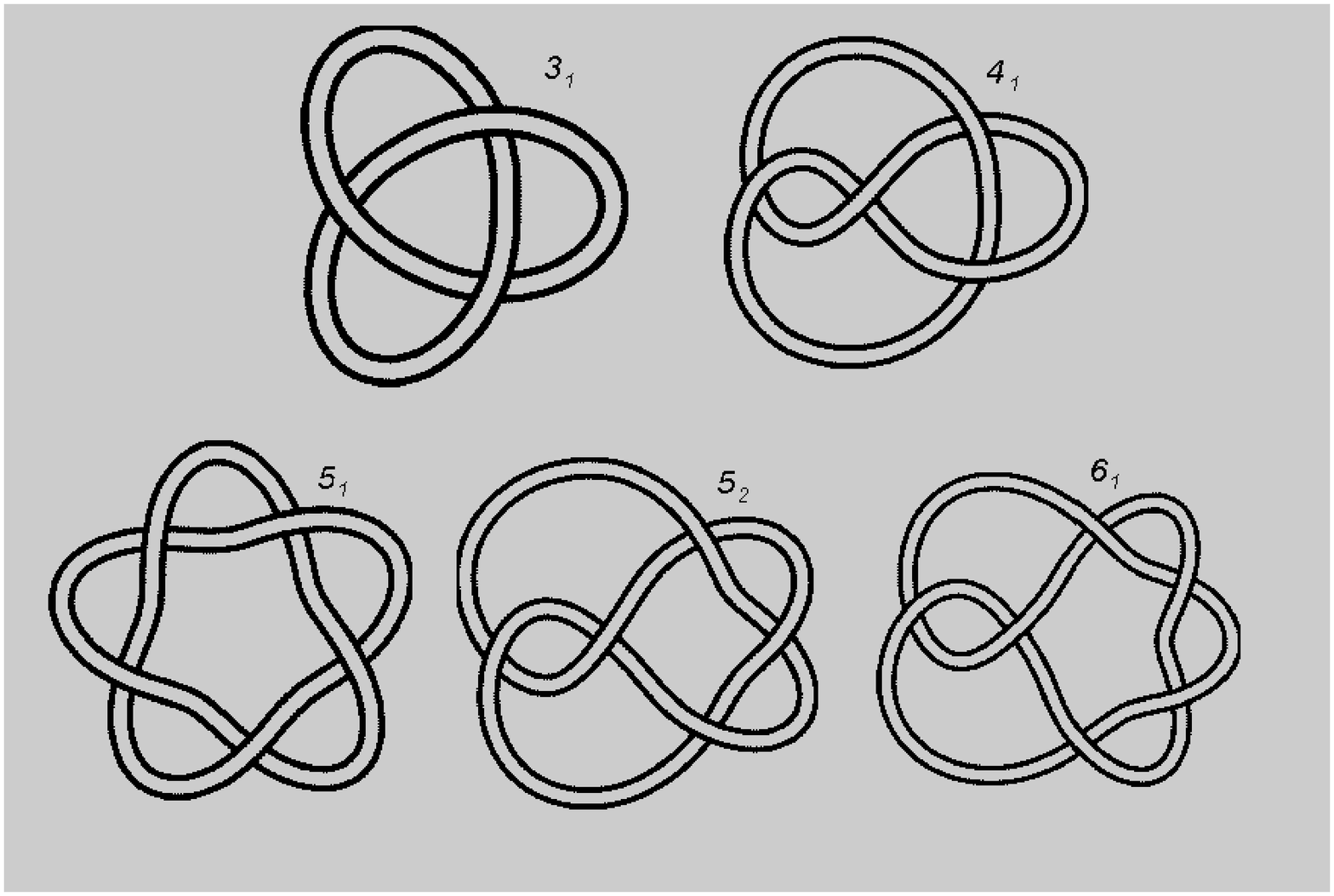}
\caption{ \label{knots}}
\end{center}
\end{figure}

In the case of non-Abelian gauge group the Wilson line operator is
more  complicated:
$$
W_{R}(c)=\textrm{tr}
P\exp\left(g\,\oint\limits_{c}A_{\mu}(x)\,dx^{\mu}\right)
$$
The vacuum expectation value has the form:
$$
<W(c)>=\sum\limits_{m=0}^{\infty}I_{m}^{R}(c)g^{2m}
$$
where the expansion are presented in terms of ordered
multidimensional integrals:
$$
I_{m}^{R}(c)=tr(R^{a_{1}}\,R^{a_{2}}...R^{a_{m}})
\int_{0}^{1}dx_{\mu_1}\int_{0}^{x_{\mu_1}}dx_{\mu_2}...\int_{0}^{x_{\mu_{m-1}}}dx_{\mu_m}
< A_{\mu_{1}}^{a_{1}}\,A_{\mu_{2}}^{a_2}...A_{\mu_{m}}^{a_m} >
$$
where $R^{a}$ are the generators of the gauge group in a
representation $R$. Using Wick theorem, and the following facts:
$$
\int_{0}^{1}dx_{\mu_1}\int_{0}^{x_{\mu_1}}dx_{\mu_2}...\int_{0}^{x_{\mu_{m-1}}}dx_{\mu_m}=\int
dx_{1}...dx_{m}\prod\limits_{k=1}^{m-1}\theta(x_{k}-x_{k+1}),
$$
$$
 <\,A_{0}^{a}(x),\,A_{\mu}(y)^{b}\,>=0, \mu=0,1,2;
$$
$$
<\,A_{\mu}^{a}(x),\,A_{\nu}^{b}(y)\,>=\dfrac{1}{2}\epsilon^{\mu\,\nu}\,\delta^{b\,c}\,\delta(x_{1}-y_{1})\,\delta(x_{2}-y_{2})\,\textrm{sign}(x_{0}-y_{0}),
\ \mu,\nu=1,2
$$
we arrive to the following formula for $I_{m}^{R}$
(J.M.F.Labastida and E.P\'erez \cite{Lab2} ):
\begin{eqnarray}
\nonumber \nonumber \label{LabasFormula}
I_{m}^{R}(c)=\sum\limits_{i_{1}<i_{2}<...<i_{m}}\epsilon_{i_1}\epsilon_{i_2}...\epsilon_{i_m}D(i_{1},i_{2},...,i_{m})+\\
\nonumber \dfrac{1}{(2!)^2}\sum\limits_{{\sigma\in S_{2}\atop
j\neq i_{1},...,i_{m-2}} \atop i_{1},...,i_{m-2}}
\epsilon_{j}^{2}\epsilon_{i_{1}},...,\epsilon_{i_{m-2}}\,D(j,\sigma,i_{1},...,i_{m-2})+\\
....
\\
\nonumber \dfrac{1}{(r!)^2}\sum\limits_{{\sigma\in S_{r}\atop
j\neq i_{1},...,i_{m-r}} \atop i_{1},...,i_{m-r}}
\epsilon_{j}^{r}\epsilon_{i_{1}},...,\epsilon_{i_{m-r}}\,D(j,\sigma,i_{1},...,i_{m-r})+
\nonumber \sum\limits_{{\sigma\in S_{m}\atop j}}
\epsilon_{j}^{m}D(j,\sigma)
\end{eqnarray}
The first term in this big sum comes from the contribution in
which all the propagators are attached to different crossings. The
second when two propagators are attached to the same crossing and
rest to different crossings and so on. The factors
$D(j,\sigma,i_{1},...,<i_{m-r})$ are group factors and they can be
computed in the following way: we attach to every crossing $i_{k}$
a group generator and $r$ generators to the crossing $j$. Then
travelling along the knot from some base point we multiply this
generators in the order that they encounter. When we arrive to the
crossing $j$ first time one encounters product of $r$ group
generators and the second time the product is rearranged in
accordance with the permutation $\sigma\in S_{r} $. After
returning to the base point we should take a trace of obtained
group factor.\\
\indent For example let us calculate the group factor
$D(3,\sigma,1,5)$, where $\sigma\in S_{2}$ (we consider the group
$S_{2}$ as a permutations of two-element set $\{b,\,c\}$ ) for the
knot projection represented~in~fig. \ref{labasgraph}. According to
the receipt, we should attach one generator $R^{a}$ to point 1,
one generator $R^{d}$ to point 5, and the product of two
generators $R^{b}R^{c}$ to point 3. Running along the knot
projection from the base point $p$ we encounter the chosen points
in the following order:  $1,\,3,\,5,\,1,\,5,\,3$, then we get the
following product of the generators attached to the points: $R^{a}
\cdot R^{b}R^{c} \cdot R^{d} \cdot R^{a} \cdot R^{d} \cdot
R^{\sigma(b)}R^{\sigma(c)}$, where we rearranged the product of
generators corresponding to the point 3 according to permutation
$\sigma$ when we arrived to the point 3 for the second time.
Finally, taking the trace we have:
$$
D(3,\sigma,1,5)\,=tr(R^{a} R^{b}R^{c} R^{d} R^{a}R^{d}
R^{\sigma(b)}R^{\sigma(c)})
$$
  \\
\\
\indent Using modern computational tools such as \verb"Maple" or
\verb"Mathematica" and formula (\ref{LabasFormula}) one can
perform computations in perturbation theory for the first several
orders.  In the table below some exact results of these calculus
for $gl(N)$-case in fundamental representation $F$ are presented:
\begin{small}
$$
\begin{array}{|c|c|c|c|c|}
\hline
  \textrm{Knot} & I_{1}^{F} & I_{2}^{F}& I_{3}^{F} & I_{4}^{F}\\
  \hline
   & & & &
    \\
  3_{1}& 3\,{N}^{2}& 3/4\,N ( 5+{N}^{2} ) &1/12\,{N}^{2} ( 53+{N}^{2})&{\frac {1}{192}}\,N ( 284+363\,{N}^{2}+{N}^{4}
  )\\
   & & & &
  \\
  4_{1} & 0 & 3\,N ( N-1 )( N+1) & 0 & {\frac {7}{144}}\,N ( N-1)( N+1 ) ( 5\,
{N}^{2}+46 )\\
 & & & &
\\
5_{1}& 5\,{N}^{2} & 5/4\,N ( 9+{N}^{2} ) & {\frac
{5}{36}}\,{N}^{2} ( 149+{N}^{2})&{\frac {5}{576}}\,N
( 1712+1287\,{N}^{2}+{N}^{4} )\\
 & & & &
\\
5_{2}&5\,{N}^{2}&1/4\,N \left( 33+17\,{N}^{2} \right)&{\frac
{5}{36}}\,{N}^{2} \left( 131+19\,{N}^{2} \right)&{\frac {1}{576}}\,N \left( 4096+10263\,{N}^{2}+641\,{N}^{4} \right)\\
  & & & &\\
  6_{1}&2\,{N}^{2}&1/2\,N \left( -13+17\,{N}^{2} \right)&1/9\,{N}^{2} \left( -47+59\,{N}^{2} \right)&{\frac {1}{288}}\,N \left( -2394+1193\,{N}^{2}+1393\,{N}^{4} \right)
  )\\
\hline
\end{array}
$$
Where we have chosen the two-dimensional projections of the knots
as in the fig. \ref{knots}
\end{small}
\\
\indent In the Abelian case $N=1$ we get the following results:
$$
\begin{array}{l}
\nonumber <\hat W_{F}(3_{1})>\,=\sum\limits_{m=0}^{\infty}
I_{m}^{F}(3_{1})\,g^{2m}=1+3\,{g}^{2}+9/2\,{g}^{4}+9/2\,{g}^{6}+{\frac
{27}{8}}\,{g}^{8}+{\frac { 81}{40}}\,{g}^{10}
+...=e^{3\,g^{2}} \\
\nonumber
< \hat W_{F}(4_{1})>\,=\sum\limits_{m=0}^{\infty} I_{m}(4_{1})\,g^{2m}=1+0\,g^2+0\,g^4+0\,g^6+0\,g^8+0\,g^{10}+...=1 \\
\nonumber <\hat W(5_{1})>\,=\sum\limits_{m=0}^{\infty}
I_{m}^{F}(5_{1})\,g^{2m}=1+5\,{g}^{2}+{\frac
{25}{2}}\,{g}^{4}+{\frac {125}{6}}\,{g}^{6}+{\frac {
625}{24}}\,{g}^{8}+{\frac {625}{24}}\,{g}^{10}+...=e^{5\,g^{2}}
 \\
\nonumber <\hat W_{F}(5_{2})>\,=\sum\limits_{m=0}^{\infty}
I^{F}_{m}(5_{2})\,g^{2m}=1+5\,{g}^{2}+{\frac
{25}{2}}\,{g}^{4}+{\frac {125}{6}}\,{g}^{6}+{\frac {
625}{24}}\,{g}^{8}+{\frac {625}{24}}\,{g}^{10}+...=e^{5\,g^{2}}
\\
\nonumber <\hat W_{F}(6_{1})>\,=\sum\limits_{m=0}^{\infty}
I^{F}_{m}(5_{3})\,g^{2m}=1+2\,{g}^{2}+2\,{g}^{4}+4/3\,{g}^{6}+2/3\,{g}^{8}+{\frac
{4}{15}}\,{g}^{ 10}
+...=e^{g^{2}} \\
\end{array}
$$
We see that in full agreement with (\ref{AbAns}) the vevs of the
Wilson loops in this cases are just products of simple exponent
operators over intersection points.  In the first non-abelian case
$N=2$ we get:
$$
\begin{array}{l}
\nonumber <\hat W_{F}(3_{1})>\,=\sum\limits_{m=0}^{\infty}
I_{m}^{F}(3_{1})\,g^{2m}=2+12\,{g}^{2}+{\frac
{27}{2}}\,{g}^{4}+19\,{g}^{6}+{\frac {73}{4}}\,{g}^ {8}+{\frac
{279}{20}}\,{g}^{10}
+...= \,? \\
\nonumber
<\hat W_{F}(4_{1})>\,=\sum\limits_{m=0}^{\infty} I_{m}^{F}(4_{1})\,g^{2m}=2+18\,{g}^{4}+{\frac {77}{4}}\,{g}^{8}+...=\,? \\
\nonumber <\hat W_{F}(5_{1})>=\sum\limits_{m=0}^{\infty}
I_{m}^{F}(5_{1})\,g^{2m}=2+20\,{g}^{2}+{\frac
{65}{2}}\,{g}^{4}+85\,{g}^{6}+{\frac {955}{8}}\,{g} ^{8}+{\frac
{871}{6}}\,{g}^{10} +...=\,?
 \\
\nonumber <\hat W_{F}(5_{2})>\,=\sum\limits_{m=0}^{\infty}
I_{m}^{F}(5_{2})\,g^{2m}= 2+20\,{g}^{2}+{\frac
{101}{2}}\,{g}^{4}+115\,{g}^{6}+{\frac {1539}{8}}\, {g}^{8}+{\frac
{1585}{6}}\,{g}^{10}
 +...=\,?
\\
\nonumber <\hat W_{F}(6_{1})>\,=\sum\limits_{m=0}^{\infty}
I_{m}^{F}(5_{3})\,g^{2m}=2+8\,{g}^{2}+55\,{g}^{4}+84\,{g}^{6}+{\frac
{4111}{24}}\,{g}^{8}+{\frac {2912}{15}}\,{g}^{10}

+...=\,? \\
\end{array}
$$

In this case it is difficult to find any regularity in the
coefficients of the expansions and sum them to the exact
expressions of vevs for kernels of associated Wilson loops
operators. It indicates that the answer for these vevs in
non-abelian case has much more complicated structure. In section 4
we present the example of such vev for the simplest trefoil knot
$3_1$ in the case $N=2$.
\subsection{Two-Component Links}
\begin{figure}[h]
\begin{center}
\includegraphics[scale=0.6,angle=-90]{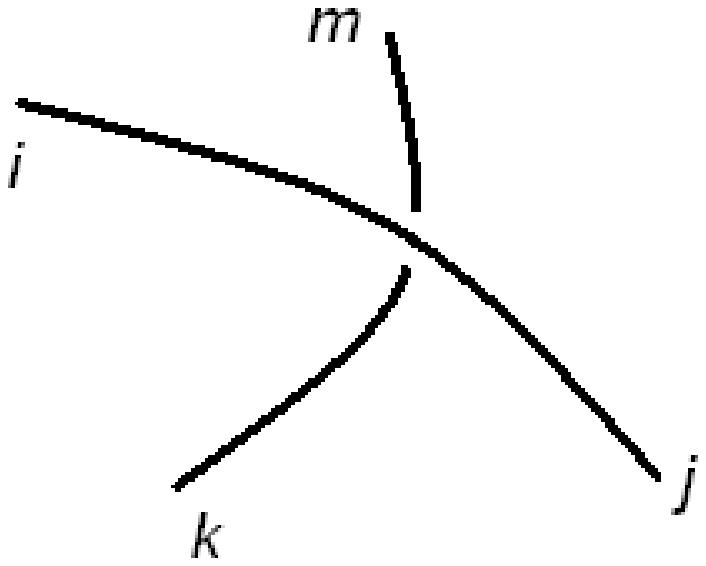}
\caption{ \label{inter}}
\end{center}
\end{figure}

In order to derive the notion of intersection point operator we
need to find a contribution to vev of Wilson loop coming from a
single two-dimensional intersection point as in fig.\ref{inter}.
For our analysis it is much more convenient to assume that the
pathes $i\,j$ and $k\,m$ belong to different contours.
\\
In this connection let us temporary proceed to the consideration
of two-component links. More precisely, we are interested in the
following quantity:
\begin{equation}
\label{cor2}
 I^{R_{1}\,R_{2}}(c_{1},c_{2})=\dfrac{<\hat W_{R_{1}}(c_{1}),\,\hat W_{R_{2}}(c_{2})>}{<\hat W_{R_{1}}(c_{1})><\hat W_{R_{2}}(c_{2})>}=\sum\limits_{k=0}^{\infty}
 I_{k}^{R_{1}\,R_{2}}(c_{1},c_{2})g^{2\,k}=1+\sum\limits_{k=1}^{\infty}
 I_{k}^{R_{1}\,R_{2}}(c_{1},c_{2})g^{2\,k}
\end{equation}
where $c_{1}$ and $c_{2}$ are two contours in $\mathbb{R}^{3}$ and
$W_{R_{i}}(c_{i})$ are associated Wilson lines operators in
representation  $R_{i}$ of the gauge group:
\begin{equation}
W_{R_{i}}(c_{i})=tr
P\exp\left(g\oint\limits_{c_{i}}A_{\mu}\,dx^{\mu}\right)
\end{equation}
In (\ref{cor2}) we divided  vev $< \hat W_{R_{1}}(c_{1}),\,\hat
W_{R_{2}}(c_{2})>$ by the product of vevs $<\hat
W_{R_{1}}(c_{1})><\hat W_{R_{2}}(c_{2})>$, which means that in
perturbation series expansion we will only take into account the
terms with different ends of the propagators attached to the
different contours, in other words, we do not take into
consideration the self-intersection points of the contours $c_{1}$
and $c_{2}$.
\subsection{ $I_{1}^{R_{1}\,R_{2}}(c_{1},c_{2})$ case}

In the order $g^2$ we have the following integral:
\begin{eqnarray}
I_{1}^{R_{1}\,R_{2}}( c_{1},\,c_{2} )=g^2\,tr(
R_{1}^{a})\,tr(R_{2}^{b} )\,\oint\limits_{c_{1}}\,dx^{\mu}
\oint\limits_{c_{2}}\,dy^{\nu} < A_{\mu}^{a}(x),\,A_{\nu}^{b}(y)
>=0\nonumber
\end{eqnarray}

\subsection{ $I_{2}^{R_{1}\,R_{2}}( c_{1}, c_{2})$ case}

In the order $g^{4}$ we have two contributions:
\begin{equation}
\label{I2} I_{2}^{R_{1}\,R_{2}}( c_{1}, c_{2})= M_{1}+M_{2}
\end{equation}
where
\begin{eqnarray}
M_{1}=tr(R_{1}^{a_{1}}R_{1}^{a_{2}})\,tr(R_{2}^{b_{1}}R_{2}^{b_{2}})
\oint\limits_{c_{1}}dx_{1}^{\mu_{1}}\int\limits_{c_{1}}^{x_{1}}dx_{2}^{\mu_{2}}
\oint\limits_{c_{2}}dy_{1}^{\nu_{1}}\int\limits_{c_{2}}^{y_{1}}dy_{2}^{\nu_{2}}
<A_{\mu_{1}}^{a_1}(x_{1}),\,A_{\nu_{1}}^{b_1}(y_{1})
>\,<A_{\mu_{2}}^{a_2}(x_{2}),\,A_{\nu_{2}}^{b_2}(y_{2}) >\\
M_{2}=tr(R_{1}^{a_{1}}R_{1}^{a_{2}})\,tr(R_{2}^{b_{1}}R_{2}^{b_{2}})
\oint\limits_{c_{1}}dx_{1}^{\mu_{1}}\int\limits_{c_{1}}^{x_{1}}dx_{2}^{\mu_{2}}
\oint\limits_{c_{2}}dy_{1}^{\nu_{1}}\int\limits_{c_{2}}^{y_{1}}dy_{2}^{\nu_{2}}
<A_{\mu_{1}}^{a_1}(x_{1}),\,A_{\nu_{2}}^{b_2}(y_{2})
>\,<A_{\mu_{2}}^{a_2}(x_{2}),\,A_{\nu_{1}}^{b_1}(y_{1}) >
\end{eqnarray}
Introducing the parameterizations $t$ and $s$ for contours $c_{1}$
and $c_{2}$ respectively we get:
$$
G(t,s)=\epsilon_{\mu\,\nu}\,\dfrac{dx^{\mu}}{dt}(t)\,\dfrac{dy^{\nu}}{ds}(s)\,\delta(x_{1}(t)-y_{1}(s))\,\delta(x_{2}(t)-y_{2}(s))
\,\dfrac{1}{2}\textrm{sign}(x_{0}(t)-y_{0}(s))=\sum\limits_{k}\epsilon_{k}\delta(t-t_{k})\delta(s-s_{k})
$$
Here $\epsilon_{k}=\pm1$ and sum runs over the intersection points
of two contours lying in the plane $(x_{1},x_{2})$ which are the
projections of three dimensional contours on the plane. Using this
notation we can rewrite the integrals for $M_{1}$ and $M_{2}$ in
the form:
\begin{eqnarray}
\label{M12}
M_{1}=tr(R_{1}^{a_{1}}R_{1}^{a_{2}})\,tr(R_{2}^{a_{1}}R_{2}^{a_{2}})\,\int\limits_{0}^{1}
dt_{1} \int\limits_{0}^{t_{1}} dt_{2}\,\int\limits_{0}^{1} ds_{1}
\int\limits_{0}^{s_{1}} ds_{2}
\,G(t_{1},\,s_{1})\,G(t_{2},\,s_{2})\\
M_{2}=tr(R_{1}^{a_{1}}R_{1}^{a_{2}})\,tr(R_{2}^{a_{2}}R_{2}^{a_{1}})\,\int\limits_{0}^{1}
dt_{1} \int\limits_{0}^{t_{1}} dt_{2}\,\int\limits_{0}^{1} ds_{1}
\int\limits_{0}^{s_{1}} ds_{2}
\,G(t_{1},\,s_{2})\,G(t_{2},\,s_{1})
\end{eqnarray}
and using the following fact:
\begin{equation}
\int\limits_{0}^{1} dt_{1} \int\limits_{0}^{t_{1}}
dt_{2}\,\int\limits_{0}^{1} ds_{1} \int\limits_{0}^{s_{1}} ds_{2}=
\int\limits_{0}^{1} dt_{1} \int\limits_{0}^{1}
dt_{2}\,\int\limits_{0}^{1} ds_{1} \int\limits_{0}^{1}
ds_{2}\,\theta(t_{1}-t_{2})\theta(s_{1}-s_{2})
\end{equation}
we have:
\begin{eqnarray}
M_{1}=tr(R_{1}^{a_{1}}R_{1}^{a_{2}})\,tr(R_{2}^{a_{1}}R_{2}^{a_{2}})\,\sum_{k_{1}\,k_{2}}\,\epsilon_{k_{1}}
\epsilon_{k_{2}}\,
\theta(t_{k_{1}}-t_{k_{2}})\,\theta(s_{k_{1}}-s_{k_{2}})\nonumber\\
\\
M_{2}=tr(R_{1}^{a_{1}}R_{1}^{a_{2}})\,tr(R_{2}^{a_{2}}R_{2}^{a_{1}})\,\sum_{k_{1}\,k_{2}}\,\epsilon_{k_{1}}\epsilon_{k_{2}}\,
\theta(t_{k_{1}}-t_{k_{2}})\,\theta(s_{k_{2}}-s_{k_{1}})\nonumber
\end{eqnarray}
and finally we arrive to the following expression:
\begin{equation}
\label{pred}
I_{2}^{R_{1}\,R_{2}}(c_{1},c_{2})=M_{1}+M_{2}=\sum_{\sigma \in
S_{2}}
tr(R_{1}^{a_{1}}R_{1}^{a_{2}})\,tr(R_{2}^{a_{\sigma(1)}}R_{2}^{a_{\sigma(2)}})\,\sum_{k_{1}\,k_{2}}\,\epsilon_{k_{1}}\epsilon_{k_{2}}\,
\theta(t_{k_{1}}-t_{k_{2}})\,\theta(s_{k_{\sigma(1)}}-s_{k_{\sigma(2)}})
\end{equation}
where $S_{2}$ is the permutation group of two elements.
\subsection{$I_{m}^{R_{1}\,R_{2}}(c_{1}\,c_{2})$ Case}

The last formula (\ref{pred}) has obvious straightforward
generalization for $I_{m}^{R_{1}\,R_{2}}(c_{1}\,c_{2})$:
\begin{equation}
 \label{Im} I_{m}^{R_{1}\,R_{2}}(c_{1}\,c_{2})=\sum\limits_{\sigma\in\,S_{m}}
M_{\sigma}
\end{equation}
where $S_{m}$ is the permutation group for a set with $m$
elements.
\begin{small}
\begin{equation}
\label{M}
M_{\sigma}=tr(R_{1}^{a_{1}}...R_{1}^{a_{m}})tr(R_{2}^{a_{\sigma(1)}}...R_{2}^{a_{\sigma(m)}})
\sum\limits_{k_{1}\,k_{2}...k_{m}}\epsilon_{k_{1}}\epsilon_{k_{2}}...\epsilon_{k_{m}}
\left(\prod\limits_{i=1}^{m-1}
\theta(t_{k_{i}}-t_{k_{i+1}})\right)\,\left(\prod\limits_{j=1}^{m-1}
\theta\left(s_{k_{\sigma(j)}}-s_{k_{\sigma(j+1)}}\right)\right)
\end{equation}
\end{small}

Let us note that in the Abelian case $M_{\sigma}$ does not contain
non-commuting lie-algebra structures and the sum of $M_{\sigma}$
is expressed through the linking number of two contours:
\begin{equation}
\sum_{\sigma\in\, S_{m}}\,M_{\sigma}\,=\sum_{\sigma\in\,
S_{m}}\left(
\sum\limits_{k_{1}\,k_{2}...k_{m}}\epsilon_{k_{1}}\epsilon_{k_{2}}...\epsilon_{k_{m}}
\left(\prod\limits_{i\,=1}^{m-1}
\theta(t_{k_{i}}-t_{k_{i+1}})\right)\,\left(\prod\limits_{j\,=1}^{m-1}
\theta\left(s_{k_{\sigma(j)}}-s_{k_{\sigma(j+1)}}\right)\right)
  \right)=\dfrac{L_{12}^{m}}{m!}
\end{equation}
where the quantity $L_{12}$ is just a sum of crossing sings over
two dimensional intersection points:
$$
L_{12}=\sum_{k}\epsilon_{k}
$$
The definition of $L_{12}$ coincides precisely with definition of
linking number of two knots. The linking number of two
two-component link known to be the topological invariant of the
link, and we arrive to the following result: in abelian case the
vev (\ref{cor2}) is just the exponent of linking number of the
link:
$$
I(c_{1},c_{2})=\sum_{m=0}^{\infty}
g^{2k}\,\dfrac{L_{12}^{m}}{m!}=\exp(g^2\,L_{12})
$$
\section{The intersection point operator method}
\subsection{The intersection point operator}
\begin{figure}[h]
\begin{center}
\includegraphics[scale=0.7,angle=-90]{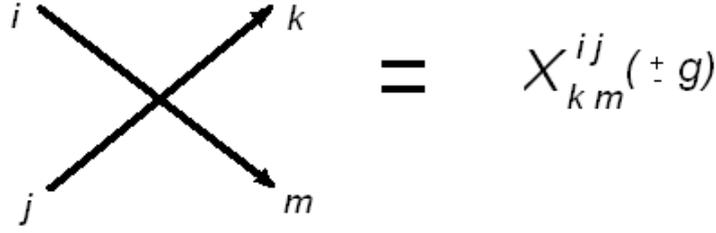}
\caption{\label{X}Representation of intersection operator $X$}
\end{center}
\end{figure}

 Using the following property of the tensor
product of operators:
$$
tr(R_{1}^{a_{1}}R_{1}^{a_{2}}...R_{1}^{a_{m}})tr(R_{2}^{a_{\sigma(1)}}R_{2}^{a_{\sigma(2)}}...R_{2}^{a_{\sigma(m)}})=
tr(R_{1}^{a_{1}}R_{1}^{a_{2}}...R_{1}^{a_{m}}\otimes
R_{2}^{a_{\sigma(1)}}R_{2}^{a_{\sigma(2)}}...R_{2}^{a_{\sigma(m)}})
$$
we can rewrite expression for $I_{m}^{R_{1}\,R_{2}}(c_{1},c_{2})$
in the form:
$$
M_{\sigma}=tr({\hat M}_{\sigma})
$$
\begin{small}
\begin{equation}
{\hat M}_{\sigma}=R_{1}^{a_{1}}...R_{1}^{a_{m}}\otimes
R_{2}^{a_{\sigma(1)}}...R_{2}^{a_{\sigma(m)}}
\sum\limits_{k_{1}\,k_{2}...k_{m}}\epsilon_{k_{1}}\epsilon_{k_{2}}...\epsilon_{k_{m}}
\left(\prod\limits_{i=1}^{m-1}
\theta(t_{k_{i}}-t_{k_{i+1}})\right)\,\left(\prod\limits_{j=1}^{m-1}
\theta\left(s_{k_{\sigma(j)}}-s_{k_{\sigma(j+1)}}\right)\right)
\end{equation}
\end{small}
Let us consider instead of the sums (\ref{cor2}) and (\ref{Im})
the following operator:
\begin{small}
\begin{eqnarray}
M=\sum_{m=0}^{\infty}\,g^{2\,m}\sum_{\sigma\in\,
S_{m}}\,R_{1}^{a_{1}}R_{1}^{a_{2}}...R_{1}^{a_{m}}\otimes
R_{2}^{a_{\sigma(1)}}R_{2}^{a_{\sigma(2)}}...R_{1}^{a_{\sigma(m)}}\sum\limits_{k_{1}\,k_{2}...k_{m}}\epsilon_{k_{1}}\epsilon_{k_{2}}...\epsilon_{k_{m}}
\prod\limits_{i=1}^{m-1} \theta(t_{k_{i}}-t_{k_{i+1}})\,
\theta\left(s_{k_{\sigma(j)}}-s_{k_{\sigma(j+1)}}\right)=\nonumber\\
=1+g^{2}\,R_{1}^{a_{1}}\otimes
R_{2}^{a_{1}}\,\sum_{k_{1}}\epsilon_{k_{1}}+g^{4}\sum_{\sigma\in
S_{2}}R_{1}^{a_{1}}R_{1}^{a_{2}}\otimes
R_{2}^{a_{\sigma(1)}}R_{2}^{a_{\sigma(2)}}\theta(t_{k_{1}}-t_{k_{2}})\theta(s_{k_{\sigma(1)}}-s_{k_{\sigma(2)}})+....
\end{eqnarray}
\end{small}
and calculate the contribution coming from one point. To find this
contribution we should to assume $k_{1}=k_{2}=...=k_{m}$ in the
$g^{2m}$-order term of $M$ expansion and take into account the
following fact:
$$
\left(\prod_{k=1}^{m-1}\theta(t_{k_{i}}-t_{k_{i+1}})\right)|_{k_{1}=k_{2}=...k_{m}}=\dfrac{1}{m!},\
\ \ \
\left(\prod_{k=1}^{m-1}\theta(s_{k_{\sigma(i)}}-s_{k_{\sigma(i+1)}})\right)|_{k_{1}=k_{2}=...k_{m}}=\dfrac{1}{m!}
$$
Let us denote:
$$
\{R_{i}^{a_{1}}R_{i}^{a_{2}}...R_{i}^{a_{m}}\}=\dfrac{1}{m!}\,\sum_{\sigma\in
S_{m}}
R_{i}^{a_{\sigma(1)}}R_{i}^{a_{\sigma(2)}}...R_{i}^{a_{\sigma(m)}}
$$
then for a single point contribution we get:
\begin{equation}
\label{Rmat}
 X(\epsilon_{j}g^{2})=
 \sum_{k=0}^{\infty}
\dfrac{\epsilon_{j}^{k}\,g^{2k}}{k!}
\{R_{1}^{a_{1}}\,R_{1}^{a_{2}}...\,R_{1}^{a_{k}}\}\otimes
\{R_{2}^{a_{1}}\,R_{2}^{a_{2}}...\,R_{2}^{a_{k}}\}
\end{equation}
 One can treat $X(\pm g^{2})$ as an operator acting in the space
$R_{1}\otimes R_{2}$. It is convenient to consider the operator
$X$ as a tensor with four indexes $X^{i\,j}_{k\,m}$, so that each
index is associated with a leg of intersection vertex fig.\ref{X}.
\\
\indent

\begin{figure}[h]
\includegraphics[scale=0.7,angle=-90]{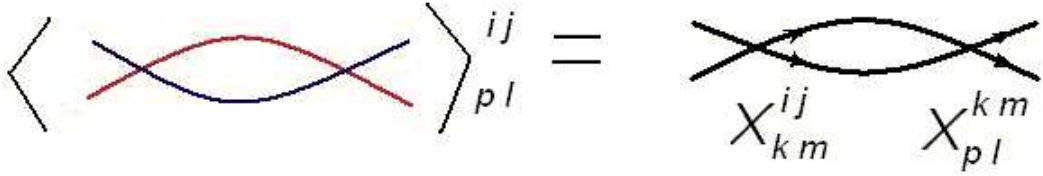}
\caption{\label{XX}The contribution coming from two adjacent
points of a link}
\end{figure}
\newpage
The main feature of this operator is that the two adjacent point
contribution can be simply expressed as corresponding product of
two such operators fig.\ref{XX}.\\
\subsection{The vev of the trefoil knot in $GL(2)$ case}
 \indent As a more explicit example let us consider how the intersection point
 operator (\ref{Rmat}) can be used for calculation of the kernels of
 polynomial invariants. Let us summate the series (\ref{Rmat})
 for operators $F^{a}$ in fundamental representation of $gl(2)$.
 In this case we have:
\begin{equation}
\label{RmatGL2} X(g^2)= X(g^2)=\sum_{k=0}^{\infty}
\dfrac{g^{2k}}{k!}
\{F^{a_{1}}\,F^{a_{2}}...\,F^{a_{k}}\}^{\otimes2}=A(g^2)\,{\hat
Id} +B(g^2)\,{\hat P}
\end{equation}
where ${\hat Id}$, ${\hat P}$ are the identity and interchange
operators in $\mathbb{C}^2\otimes\mathbb{C}^2$ and:
$$
A(g^2)=1/6\,{e^{g^2}}+1/6\,{e^{g^2}}g^2+5/6-1/3\,g^2,
$$
$$
B(g^2)=2/3\,{e^{g^2}}-2/3+1/6\,g^2+1/6\,{e^{g^2}}g^2
$$
$$
{\hat Id}=\left[ \begin {array}{cccc}
1&0&0&0\\\noalign{\medskip}0&1&0&0\\\noalign{\medskip}0&0&1&0\\\noalign{\medskip}0&0&0&1\end
{array}
 \right],\ \ \ \
 {\hat P}=\left[ \begin {array}{cccc} 1&0&0&0\\\noalign{\medskip}0&0&1&0\\
 \noalign{\medskip}0&1&0&0\\\noalign{\medskip}0&0&0&1\end {array}
 \right]
$$
\begin{figure}[h]
\begin{center}
\includegraphics[scale=0.5,angle=-90]{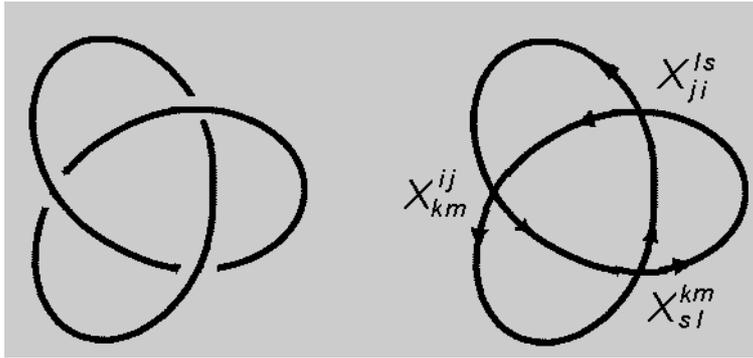}
\caption{\label{31}Two dimensional projection of the trefoil knot
and contraction of intersection operators.}
\end{center}
\end{figure}
To find the kernel of polynomial invariant, for example of the
trefoil knot, we need to chose some projection of the knot to
two-dimensional plane. For instance as in the fig.\ref{31}. Then,
we attach to every intersection point of obtained two-dimensional
curve the tensor $X(\epsilon g^2)$, where $\epsilon$ is the sign
of the intersection point defined by~(\ref{signs}). Finally, to
find the kernel of polynomial invariant in this case, we contract
the indexes of the tensors in accordance with two dimensional
diagram fig.\ref{31}:
$$
<\hat W(3_1)>=\sum\limits_{i\,j\,\,k\,m\,s\,l}
X^{i\,j}_{k\,m}\,X^{k\,m}_{s\,l}\,X^{l\,s}_{j\,k}=
$$
$$
2\, ( 5/6\,{e^{{g}^{2}}}+1/3\,{e^{{g}^{2}}}{g}^{2}+1/6-1/6\,{g}^{
2} ) ^{3}+
$$
\begin{equation}
\label{tref} +6\,( 1/6\,{e^{{g}^{2}}}+1/6\,{e^{{g}^{2}}}{g}^{2
}+5/6-1/3\,{g}^{2} ) ^{2} ( 2/3\,{e^{{g}^{2}}}-2/3+1/6\,{g}
^{2}+1/6\,{e^{{g}^{2}}}{g}^{2} )+
\end{equation}
$$
+ 2\, ( 2/3\,{e^{{g}^{2}}}-
2/3+1/6\,{g}^{2}+1/6\,{e^{{g}^{2}}}{g}^{2} ) ^{3}
$$
The expansion of $<\hat W(3_1)>$ in $g^2$ coincide precisely with
the result obtained in section $3.1$ for the trefoil knot $3_1$ in
the $N=2$ case by means of Labastida-P\'erez formula:
$$
<\hat W(3_1)>=2+12\,g^2+{\frac {27}{2}}{g}^{4}+19\,{g}^{6}+{\frac
{73}{4}}{g}^{8}+{ \frac {279}{20}}{g}^{10}+O ( {g}^{12} )
$$
Therefore, we observed an interesting property of CS theory in the
temporal gauge with the propagator (\ref{Prop}): the vevs of
Wilson loops are factorized into the product of the intersection
point operators corresponding to the crossings of two-dimensional
projection of the knot. We note that in this case the vevs of
Wilson loops are not knot invariants and depends on the projection
chosen, moreover, the kernels of the polynomial invariants are not
polynomials in $e^{g^2}$ anymore (as it could be seen from (\ref{tref})).\\
\indent
 Of course, without the prescription depending term in the
propagator (\ref{Green}) the theory is incomplete, and appropriate
choice for this term is needed. Nevertheless, the first term of
the propagator (\ref{Green}) that we used in this work, contains
only information about crossings, and the second, prescription
depending term, does not depend on the crossing sings, as it does
not depend on $x_{0}$. This leads us to the conjecture, that in
the presence of the prescription depending term, the property of
vevs to be factorized into the product of some tensors
corresponding to crossings, should be conserved.
\section{Conclusion}
There are a lot of the combinatorial constructions for the knot
polynomial invariants in terms of regular two-dimensional
projections. For example the Jones polynomial arising from the
braid group representations \cite{Guad2} or the Kauffman
construction of Jones polynomials in terms of $R$-matrices and the
"creation-annihilation" operators \cite{Kauff2}. All this
constructions provide some tensors corresponding to intersection
points, and some additional tensors corresponding to free lines,
like the Kauffman "creation"  and "annihilation" operators which
correspond to critical points of the knots in the Morse
representation.  The natural way for deriving these
representations form CS theory is to use the temporal gauge fixing
as the perturbation theory in this gauge depends only on
two-dimensional representation of the knot. The CS theory in the
temporal gauge with propagator (\ref{Prop}) contains only
information about crossings, and operators corresponding to the
Kauffman creation annihilation operators can not be derived by
means of this propagator. In this way, we should conclude, that
the prescription depending term in (\ref{Green}) plays a crucial
role in the construction of correct perturbation theory for CS in
the temporal gauge.\\
\indent  To find exact expression for this term we need some
additional physical restrictions on the form of the propagator. As
an example of such a restriction we can demand that the propagator
gives the perturbative series expansion for Wilson lines is in
agreement with some general properties of CS theory, for example
the factorization theorem \cite{Lab1}-\cite{Lab3}. The examples of
restrictions on prescription depending term arising from
factorization theorem can be found in \cite{Lab2}. Another example
of restriction gives consideration of the unknot vevs for a
projection without intersection points. In this case the first
term of the propagator (\ref{Green}) does not play any role and
the perturbation series expansion contains only multidimensional
integrals on products of prescription depending terms. We should
demand that this series expansion coincides with the known series
for vevs of unknot. All this will be considered elsewhere.
\begin{Large}\\
\\
\\
\textbf{Acknowledgments} \end{Large}\\
\\
The author is grateful to A.Morozov for fruitful discussions and
interest to this work. The work was partly supported by RFBR grant
09-02-00393, RFBR grant 06-01-92054-$KE_{a}$, RFBR-CNRS grant
09-01-93106,  and grant for support of scientific schools
NSh-3036.2008.2. The work was
also supported in part by the "Dynasty" Foundation.  \\
\\

\end{document}